\documentclass[12pt]{iopart}
\pdfoutput=1
\usepackage{bbm}
\usepackage{graphicx}
\usepackage{xcolor}
\usepackage{cite}

\begin{document}
\title[XYZ Richardson-Gaudin models]{Integrable spin-$\frac{1}{2}$ Richardson-Gaudin XYZ models in an arbitrary magnetic field}
\author{Pieter W. Claeys$^{1,2}$, Claude Dimo$^3$, Stijn De Baerdemacker$^1$, Alexandre Faribault$^3$}
\address{ 
$^1$ Department of Physics and Astronomy, Ghent University, Krijgslaan 281-S9, 9000 Ghent, Belgium \\
$^2$ Department of Physics, Boston University, 590 Commonwealth Ave., Boston, MA 02215, USA \\
$^3$ Universit\'e de Lorraine, CNRS, LPCT, F-54000 Nancy, France }
\ead{pwclaeys@bu.edu, alexandre.faribault@univ-lorraine.fr}

\begin{abstract}
We establish the most general class of spin-$\frac{1}{2}$ integrable Richardson-Gaudin models including an arbitrary magnetic field, returning a fully anisotropic (XYZ) model. The restriction to spin-$\frac{1}{2}$ relaxes the usual integrability constraints, allowing for a general solution where the couplings between spins lack the usual antisymmetric properties of Richardson-Gaudin models. The full set of conserved charges are constructed explicitly and shown to satisfy a set of quadratic equations, allowing for the numerical treatment of a fully anisotropic central spin in an external magnetic field. While this approach does not provide expressions for the exact eigenstates, it allows their eigenvalues to be obtained, and expectation values of local observables can then be calculated from the Hellmann-Feynman theorem.

\end{abstract}

\date{\today}

%
%

\section{Introduction}

Gaudin models are a specific class of quantum integrable models characterised by a large set of mutually commuting conserved charges \cite{gaudin_1976,gaudin_bethe_2014,dukelsky_colloquium_2004,ortiz_exactly-solvable_2005}. Typically, each conserved charge contains interaction terms between a single (central) spin and the full set of other (bath) spins in the system. Such interactions can either be fully isotropic as in XXX models or fully anisotropic as in XYZ models, while the intermediate XXZ models maintain $U(1)$-rotation symmetry in the XY-plane but show anisotropy along the, now distinct, $z$-direction. Remarkably, the XXX and XXZ models can be extended by adding a magnetic field along the $z$-direction, leading to Richardson-Gaudin (RG) models \cite{dukelsky_colloquium_2004,ortiz_exactly-solvable_2005}. 

In this work, we address the question of the most general spin-$\frac{1}{2}$ Richardson-Gaudin model which remains integrable in the presence of an arbitrary external magnetic field. Crucially, the demand that all interactions are fully antisymmetric can be removed for spin-$\frac{1}{2}$ models, allowing for non-antisymmetric integrable models whose integrability supports an external magnetic field. This work builds on the recent realisation that the XXZ model can also be extended with an arbitrary magnetic field (breaking the common $U(1)$ symmetry of XXZ models) through the restriction to spin-$\frac{1}{2}$ models \cite{lukyanenko_boundaries_2014,lukyanenko_integrable_2016,claeys_readgreen_2016,links_solution_2017,faribault_field_2017}. 

Here, we go a step beyond and present a general parametrisation of integrable spin-$\frac{1}{2}$ XYZ models in an external field. While their definition initially assumes that the field is non-zero, the field components can all be individually set to zero through proper limits. The fact that an external magnetic field does not break their integrability could allow these models to provide a more realistic description of the decoherence of a central spin, in a field, coupled to a nuclear spin bath. Indeed, when e.g. quadrupole coupling is included, the appropriate exchange interactions become fully anisotropic (XYZ) \cite{sinitsyn_quad_2012}.

The approach used in this work allows for the explicit construction of the conserved charges defining a class of XYZ integrable models. In a series of papers, Skrypnyk has investigated the consequences of a non-skew-symmetric classical $r$-matrix satisfying the classical Yang-Baxter equation on the construction of more extensive Richardson-Gaudin models\cite{skrypnyk_integrable_2006,skrypnyk_generalized_2007,skrypnyk_generalized_2007_1,skrypnyk_non-skew-symmetric_2009,skrypnyk_non-skew-symmetric_2009-1,skrypnyk_elliptic_2012,skrypnyk_generalized_2014,skrypnyk_twisted_2016,skrypnyk_z_2016,skrypnyk_separation_2017}. We here show how our results are equivalent to Skrypnyk's construction, with the main advantage being that our method guarantees generality within the class of proposed interactions and starts from the more straightforward demand that a set of proposed operators commute, circumventing the more involved Yang-Baxter approach.

The resulting models fall within the general class of RG models for which it was shown that their conserved charges satisfy a set of quadratic relations which can be efficiently solved to find the eigenvalues of the conserved charges directly \cite{claeys_readgreen_2016,faribault_field_2017,babelon_bethe_2007,faribault_xxxsolve_2011,elaraby_xxxsolve_2012,claeys_xxzsolve_2015,links_completeness_2017,dimo_quadratic_2018}. These quadratic relations are given here and it is shown how expectation values of local operators immediately follow from the Hellmann-Feynman theorem.  While this does not provide a systematic way to explicitly construct eigenstates, our capacity to easily solve for eigenvalues is sufficient to allow for extensive numerical studies of these models, following similar ones for the elliptic model \cite{esebbag_elliptic_2015,relano_excited-state_2016}.

In the following, we recapitulate the different classes of antisymmetric RG models, including the elliptic XYZ model, in Section \ref{sec:RGmodels} and continue in Section \ref{sec:spin12RGmodels} with the general parametrisation of spin-$\frac{1}{2}$ XYZ models in arbitrary magnetic fields. Section \ref{sec:solution} then details the quadratic relations among the conserved charges and the use of the Hellmann-Feynman theorem, combined with numerical results, while Section \ref{sec:concl} is reserved for conclusions.

\section{Richardson-Gaudin models}
\label{sec:RGmodels}

\emph{Gaudin models.} -- Gaudin originally presupposed a set of operators, acting as conserved charges, of the form \cite{gaudin_1976, gaudin_bethe_2014}
\begin{equation}
Q_i = \sum_{j \neq i}^L \left[\Gamma^x_{ij} S^x_i S^x_j + \Gamma^y_{ij} S^y_i S^y_j+\Gamma^z_{ij} S^z_i S^z_j \right], \qquad i=1 \dots L,
\end{equation}
where the $S^\alpha_i, \forall i=1,2,\dots L$ are $L$ distinct realisations of the $su(2)$ spin algebra satisfying $[S_i^{\alpha},S_j^{\beta}] = i \sum_{\gamma} \epsilon_{\alpha \beta \gamma} \delta_{ij}S_i^{\gamma}$. In order for these operators to behave as conserved charges, they should mutually commute $[Q_i,Q_j]=0, \forall i,j=1 \dots L$. This demand leads to the so-called \emph{Gaudin equations}
\begin{eqnarray}
&\Gamma^{\alpha}_{ij} = -\Gamma^{\alpha}_{ji}, \qquad &\forall i \neq j, \nonumber\\
&\Gamma^{\alpha}_{ij}\Gamma^{\beta}_{jk}+\Gamma^{\beta}_{ki}\Gamma^{\gamma}_{ij}+\Gamma^{\gamma}_{jk}\Gamma^{\alpha}_{ki} =0, \qquad &\forall i \neq j \neq k,
\end{eqnarray}
for any permutation $(\alpha, \beta, \gamma)$ of $(x,y,z)$. Different solutions to these equations are known, and they all define distinct classes of integrable Gaudin models \cite{gaudin_1976,gaudin_bethe_2014}. All of these solutions express the couplings $\Gamma^{\alpha}_{ij}$ as antisymmetric functions of a set of real variables (so-called inhomogeneities) $\{\epsilon_1, \dots, \epsilon_L\}$.
\begin{itemize}
\item The XXX (rational) model
\begin{eqnarray}\label{RG:XXXsym}
\Gamma^x_{ij}=\Gamma^y_{ij}=\Gamma^z_{ij}=\frac{1}{\epsilon_i-\epsilon_j}
\end{eqnarray}
\item The XXZ (hyperbolic) model
\begin{eqnarray}\label{RG:XXZsym}
\Gamma^x_{ij}=\Gamma^y_{ij}= \frac{1}{\sinh(\epsilon_i-\epsilon_j)}, \nonumber\\
\Gamma^z_{ij}= \coth(\epsilon_i-\epsilon_j)
\end{eqnarray}
\item The XYZ (elliptic) model
\begin{eqnarray}\label{RG:XYZsym}
\Gamma^x_{ij}=\frac{1+k \ \mathrm{sn}^2(\epsilon_i-\epsilon_j,k)}{\mathrm{sn}(\epsilon_i-\epsilon_j,k)}, \nonumber\\
\Gamma^y_{ij}= \frac{1-k \  \mathrm{sn}^2(\epsilon_i-\epsilon_j,k)}{\mathrm{sn}(\epsilon_i-\epsilon_j,k)}, \nonumber\\
\Gamma^z_{ij}=\frac{\mathrm{cn}(\epsilon_i-\epsilon_j,k)\mathrm{dn}(\epsilon_i-\epsilon_j,k)}{\mathrm{sn}(\epsilon_i-\epsilon_j,k)}
\end{eqnarray}
\end{itemize}
Here, the XYZ model is expressed in terms of Jacobi elliptic functions with arbitrary elliptic modulus $k$. It can easily be shown that the XXX model can be obtained as a particular limit of the XXZ model, which is in turn a specific limit of the (elliptic) XYZ model. The exact eigenvalues and eigenstates of these elliptic models were obtained by Sklyanin and Takebe using the quasi-classical limit of the corresponding inhomogeneous eight-vertex model \cite{sklyanin_xyz_2016}. Due to the lack of $U(1)$ symmetry the eigenstates are not split into conserved magnetisation $S^z = \sum_{i=1}^L S_i^z$ sectors and the resulting construction is much more convoluted than the usual algebraic Bethe ansatz for eigenstates of the XXX and XXZ models \cite{ortiz_exactly-solvable_2005}.

\emph{Richardson-Gaudin models.} -- Following the original derivation of the Gaudin equations, it was realised that the conserved charges could be extended by including a single term corresponding to a magnetic field along one of the  $(xyz)$-directions\footnote{The choice of $z$-direction for the magnetic field is arbitrary and can be seen as a gauge freedom.}\cite{dukelsky_class_2001,amico_integrable_2001}
\begin{equation}
Q_i = B^z_i S_i^z + \sum_{j \neq i}^L \left[\Gamma^x_{ij} S^x_i S^x_j + \Gamma^y_{ij} S^y_i S^y_j+\Gamma^z_{ij} S^z_i S^z_j \right], \qquad i=1 \dots L.
\end{equation}
Demanding these operators to commute then leads to the condition $B^z_i = B^z_j, \forall i,j$, while restricting the Gaudin equations to the XXZ equations with $\Gamma^x_{ij}=\Gamma^y_{ij}$, 
\begin{eqnarray}
&\Gamma^x_{ij} = -\Gamma^x_{ji}, \qquad \Gamma^z_{ij} = -\Gamma^z_{ji}, \qquad &\forall i \neq j,\nonumber\\
&\Gamma^x_{ij}\Gamma^x_{jk}-\Gamma^x_{ik}\left(\Gamma^z_{ij}+\Gamma^z_{jk}\right)=0, \qquad &\forall i \neq j \neq k.
\end{eqnarray}
As such, the presence of this additional term does not break the integrability of the XXX and XXZ models, but no longer allows for a fully anisotropic XYZ model. Any non-zero magnetic field effectively enforces isotropy in the plane orthogonal to its orientation. These models were then termed Richardson-Gaudin models, since the resulting operators correspond to the conserved charges of the reduced Bardeen-Cooper-Schrieffer model as solved by Richardson \cite{richardson_restricted_1963,richardson_exact_1964,cambiaggio_integrability_1997}. The exact solution of these models by Bethe ansatz was obtained independently by Richardson and Gaudin, where the main result is that the restriction to XXZ models allows for clear generalised spin raising/lowering operators which can be used to construct exact eigenstates \cite{ortiz_exactly-solvable_2005}.

\section{Spin-$\frac{1}{2}$ XYZ Richardson-Gaudin models} 
\label{sec:spin12RGmodels}

In the previous section, all commutation relations were obtained using only the $su(2)$ algebra. As such, the resulting integrable models hold for arbitrary spin-$s$ representations of the different spins in the systems. However, by restricting to the spin-$\frac{1}{2}$ representation it is possible to relax the integrability constraints through the use of the additional relation $S_i^{\alpha}S_i^{\beta}=\frac{i}{2}\sum_{\gamma}\epsilon_{\alpha\beta\gamma}S_i^{\gamma}+\frac{1}{4}\delta_{\alpha\beta} \mathbbm{1}$.

In the following, we will consider conserved charges of the form
\begin{equation}
Q_i = B^x_i S^x_i + B^y_i S^y_i + B^z_i S^z_i  + \sum_{j\neq i}^L \left[\Gamma^x_{ij} S^x_i S^x_j + \Gamma^y_{ij} S^y_i S^y_j + \Gamma^z_{ij} S^z_i S^z_j \right],
\label{genqi}
\end{equation}
where the $S^{\alpha}_i,\forall i=1,2,\dots L$ are now $L$ distinct spin-$\frac{1}{2}$ realisations of the $su(2)$ spin algebra. Requiring that these $L$ operators commute $[Q_i,Q_j]=0, \forall i,j$ again leads to constraints on the allowed values of the coupling parameters. It is shown in \ref{app:commutators} that the following relations, for any permutation $(\alpha, \beta, \gamma)$ of $(x,y,z)$,
\begin{eqnarray}\label{eq:spin12constraints}
&\Gamma^{\beta}_{i j} B^\alpha_j  + \Gamma^{ \gamma}_{j i} B^\alpha_i = 0,  \qquad &\forall  i \neq j, \nonumber\\
&\Gamma^{\alpha}_{ik}\Gamma^{\beta}_{jk} - \Gamma^{\alpha}_{ij}\Gamma^{\gamma}_{jk} - \Gamma^{\beta}_{ji}\Gamma^{\gamma}_{ik} = 0, \qquad &\forall i \neq j \neq k,
\end{eqnarray}
are sufficient to ensure mutual commutation of the $Q_i$ operators and therefore, integrability. Crucially, the antisymmetry condition is removed by the restriction to spin-$\frac{1}{2}$ models, which allows for a non-zero solution to the additional constraints on the magnetic field. For arbitrary spin, it can be checked how the antisymmetry condition combined with a nonzero $B_i^z$ imposes $\Gamma^x_{ij} = \Gamma^y_{ij}, \forall i,j$. 

We now make the crucial assumption that a non-zero magnetic field component is present in the model (more specifically, we assume that the magnetic field has a nonzero component along the $z$-axis in order to make the connection with previous works). Since the commutation relations will not be affected by a rescaling of the $Q_i$ operators, we can then set $B^z_i=1, \forall i$ without any loss of generality. As shown in \ref{app:constraints}, the exact and unique solution to the integrability constraints (\ref{eq:spin12constraints}) can be parametrised as
\begin{eqnarray}\label{spin12:XYZmodels}
B^x_i = \frac{\gamma}{\sqrt{\alpha_x \epsilon_i+\beta_x}},\qquad &&\Gamma^x_{ij} = g\frac{\sqrt{(\alpha_x \epsilon_i +\beta_x)(\alpha_y \epsilon_j +\beta_y)}}{\epsilon_i-\epsilon_j} \nonumber\\
B^y_i = \frac{\lambda}{\sqrt{\alpha_y \epsilon_i+\beta_y}}, \qquad &&\Gamma^y_{ij} = g\frac{\sqrt{(\alpha_y \epsilon_i + \beta_y)(\alpha_x \epsilon_j +\beta_x)}}{\epsilon_i-\epsilon_j} \nonumber\\
B^z_i = 1, \qquad &&\Gamma^z_{ij} =g \frac{\sqrt{(\alpha_x \epsilon_j + \beta_x)(\alpha_y \epsilon_j +\beta_y)}}{\epsilon_i-\epsilon_j} 
\end{eqnarray}

While this is expressed in terms of 7 free variables $\{\alpha_x,\beta_x,\alpha_y,\beta_y,\gamma,\lambda,g\}$ and the usual set of $L$ inhomogeneities $\{\epsilon_1, \dots, \epsilon_L\}$ it should be noted that $g$, $\gamma$ and $\lambda$ can be absorbed in the other variables through a common rescaling. However, this choice of parametrisation best makes clear the freedom offered by the current model, where the prefactor of each term in the conserved charges can be chosen freely.

\emph{Connection to non-skew-symmetric matrices.} -- Such models were also obtained by Skrypnyk in the study of non-skew-symmetric $r$-matrices, further cementing the connection between the presence of an arbitrary magnetic field and lack of antisymmetry in the interactions \cite{skrypnyk_elliptic_2012}. Generally, the antisymmetric models presented in Eqs.~(\ref{RG:XXXsym}-\ref{RG:XYZsym}) can be derived starting from a skew-symmetric matrix satisfying the classical Yang-Baxter equations. In a series of works by Skrypnyk, it was shown how non-skew-symmetric matrices could be used to construct a similar but distinct class of RG models, which also lacked the antisymmetry in their interactions and allowed for the presence of arbitrary magnetic fields \cite{skrypnyk_integrable_2006,skrypnyk_generalized_2007,skrypnyk_generalized_2007_1,skrypnyk_non-skew-symmetric_2009,skrypnyk_non-skew-symmetric_2009-1,skrypnyk_elliptic_2012,skrypnyk_generalized_2014,skrypnyk_twisted_2016,skrypnyk_z_2016,skrypnyk_separation_2017}. While the solution (\ref{spin12:XYZmodels}) cannot be mapped to an antisymmetric elliptic model, it can be mapped to a non-skew-symmetric one, as shown in Ref.~\cite{skrypnyk_elliptic_2012}.  However, for arbitrary spin these models necessitate additional terms $S_i^{\alpha}S_i^{\alpha}$ and fall outside of the parametrisation proposed in Eq.~(\ref{genqi}). The spin-$\frac{1}{2}$ demand hence obscures the distinction between these models as the additional terms reduce to constants. Furthermore, while the XXX and XXZ models in arbitrary fields could be mapped to the antisymmetric models of Eqs.~(\ref{RG:XXXsym}-\ref{RG:XXZsym}) in the limit of a vanishing external field, this was done by making use of the additional $U(1)$ symmetry lacking in the XYZ model \cite{links_solution_2017}. As such, the fact that the presented XYZ model can not be mapped to an elliptic antisymmetric model reinforces how the non-skew-symmetric models differ from the symmetric ones, irrespective of the spin-$\frac{1}{2}$ restriction.

\emph{Specific limits and symmetries.} -- Taking $\beta_x=\beta_y=0$ returns the spin-$\frac{1}{2}$ XXZ model introduced by Lukyanenko \emph{et al.} \cite{lukyanenko_integrable_2016}, whereas taking $\alpha_x=\alpha_y=0$ returns the regular XXX model in an arbitrary magnetic field. In these cases, when $\gamma=\lambda=0$ also holds, this results in total spin projection (along the $z$-axis) $U(1)$ symmetry. In the XXZ case ($\beta_x=\beta_y=0$) with $\gamma=\lambda=0$, this then also leads to the less-common Read-Green symmetry of RG models \cite{ibanez_exactly_2009,rombouts_quantum_2010,links_exact_2015}. Generally, if $\beta_x/\alpha_x=\beta_y/\alpha_y$ and $\gamma=\lambda$, we also have a reflection symmetry along the $x=y$ line (exchanging $x$ and $y$ leaves all conserved charges invariant). For arbitrary $\alpha_x, \alpha_y, \beta_x, \beta_y$ with $\gamma=\lambda=0$, the $U(1)$ symmetry gets broken down to a discrete parity symmetry, where the spectrum can be divided into odd- and even-parity sectors similar to the elliptic XYZ model \cite{esebbag_elliptic_2015}. 

It should also be mentioned that, while the initial assumption was that a non-zero magnetic field component was present, this constraint can be removed from the final model by rescaling the conserved charges with $g^{-1}$ and taking the limit $g\to \infty$\footnote{This removes the field along the $z$-direction, where fields along the $x$- and $y$- directions can be recovered by scaling $\gamma$ and $\lambda$ with $g$.}. As such, when no magnetic field is present and $\alpha_x=\alpha_y=0$, these systems have the total spin $SU(2)$ symmetry of XXX Gaudin models \cite{gaudin_bethe_2014}.

\section{Solution method}
\label{sec:solution}

\subsection{Eigenvalues of the conserved charges}

The construction outlined in the previous sections gives rise to a generic set of conserved charges of the form 
\begin{eqnarray}
&&Q_i = \left(S_i^z+\frac{1}{2}\right)+\frac{\gamma}{\sqrt{\alpha_x\epsilon_i+\beta_x}}S_i^x+\frac{\lambda}{\sqrt{\alpha_y\epsilon_i+\beta_y}}S_i^y  \nonumber\\
&&+g\sum_{j \neq i}^L \frac{1}{\epsilon_i-\epsilon_j}\left[\sqrt{\alpha_x\epsilon_i+\beta_x}\sqrt{\alpha_y\epsilon_j+\beta_y}S_i^x S_j^x + \sqrt{\alpha_y\epsilon_i+\beta_y}\sqrt{\alpha_x\epsilon_j+\beta_x}S_i^y S_j^y\right] \nonumber\\
&&+g \sum_{j \neq i}^L \frac{\sqrt{\alpha_x\epsilon_j+\beta_x}\sqrt{\alpha_y\epsilon_j+\beta_y}}{\epsilon_i-\epsilon_j}\left(S_i^z S_j^z-\frac{1}{4}\right), 
\end{eqnarray}
where a constant, which obviously does not affect integrability, has been added to (\ref{genqi}) simply to make the following results more clear. Recently, it has been realised that it is possible to obtain the eigenvalues of the conserved charges $Q_i$ of spin-$\frac{1}{2}$ RG models without explicit construction of the Bethe eigenstates. Starting from the set of quadratic equations\cite{claeys_readgreen_2016,faribault_field_2017,babelon_bethe_2007,faribault_xxxsolve_2011,elaraby_xxxsolve_2012,claeys_xxzsolve_2015,links_completeness_2017,dimo_quadratic_2018} \footnote{While it was not explicitly mentioned in Ref. \cite{dimo_quadratic_2018} the validity of the quadratic equations requires non-zero field components (whose limits can afterwards be taken to zero if needed) and therefore does not apply to the traditional elliptic XYZ models. However, this provides the quadratic relations for our proposed XYZ parametrisation.},
\begin{eqnarray}
Q_i^2 =&& Q_i+\frac{1}{4}\left(\frac{\gamma^2}{\alpha_x\epsilon_i+\beta_x}+\frac{\lambda^2}{\alpha_y\epsilon_i+\beta_y}\right)\nonumber \\
&&-\frac{g}{2}\sum_{j \neq i}^L\sqrt{\alpha_x\epsilon_j+\beta_x}\sqrt{\alpha_y\epsilon_j+\beta_y}\left(\frac{Q_i-Q_j}{\epsilon_i-\epsilon_j}\right) \nonumber \\
&&+\frac{g^2}{16}\sum_{j \neq i}^L\left[\frac{\sqrt{\alpha_x\epsilon_i+\beta_x}\sqrt{\alpha_y\epsilon_j+\beta_y}-\sqrt{\alpha_y\epsilon_i+\beta_y}\sqrt{\alpha_x\epsilon_j+\beta_x}}{\epsilon_i-\epsilon_j}\right]^2,
\end{eqnarray}
the observation that all conserved charges commute, and therefore share a common eigenbasis, implies that the set of eigenvalues $\{q_1, \dots, q_L\}$ corresponding to a given eigenstate necessarily satisfy the same set of quadratic equations, which can be efficiently solved numerically. Individual single states can be targeted by continuously deforming a given set of eigenvalues from the trivial non-interacting limit to the interacting model we wish to study \cite{faribault_xxxsolve_2011, elaraby_xxxsolve_2012,claeys_xxzsolve_2015}. Such methods have already proven to be extremely powerful in the study of RG models (see e.g. \cite{faribault_cs_2013,faribault_cs2_2013,van_den_berg_competing_2014,claeys_floquet_2018,tschirhart_dicke_2018,rubio_benchmarking_2018}), and were made especially simple after it was shown that overlaps with the exact eigenstates can be efficiently expressed, as determinants, directly in terms of these eigenvalues \cite{faribault_det_2012,tschirhart_det_2014,claeys_xxzsolve_2015,claeys_det_2015,faribault_field_2017,claeys_inner_2017}. However, for the general XYZ models discussed in this work, it remains an open question whether such determinant representations can be built. Finally, let us remark that in the thermodynamic limit of infinite system sizes, these equations reduce to an integral equation which is expected to remain tractable using the methods presented in Ref. \cite{shen_ground-state_2018} for XXZ RG models.

\subsection{Local observables}

Even without the exact Bethe eigenstates, expectation values of local spin observables can be calculated from the Hellmann-Feynman theorem since the model remains integrable for arbitrary magnetic fields. Given the set of eigenvalues $\{q_1, \dots, q_L\}$ defining a given eigenstate, the expectation values in that given eigenstate follow as
\begin{eqnarray}
\langle S_i^x \rangle = \sqrt{\alpha_x\epsilon_i+\beta_x}\frac{\partial q_i}{\partial \gamma}, \qquad \langle S_i^y \rangle = \sqrt{\alpha_y\epsilon_i+\beta_y}\frac{\partial q_i}{\partial \lambda}, \nonumber \\
\langle S_i^z \rangle = q_i-g\frac{\partial q_i}{\partial g} - \gamma \frac{\partial q_i}{\partial \gamma} - \lambda \frac{\partial q_i}{\partial \lambda}.
\end{eqnarray}
These partial derivatives can either be calculated numerically through a finite-difference method, or by solving a linear set of equations. Since the eigenvalues obey a set of quadratic equations, their derivatives will indeed obey a linear system which is straightforward to compute and which can then easily be solved numerically.

\subsection{Numerical results}

\begin{figure}[ht]
\includegraphics[width=\textwidth]{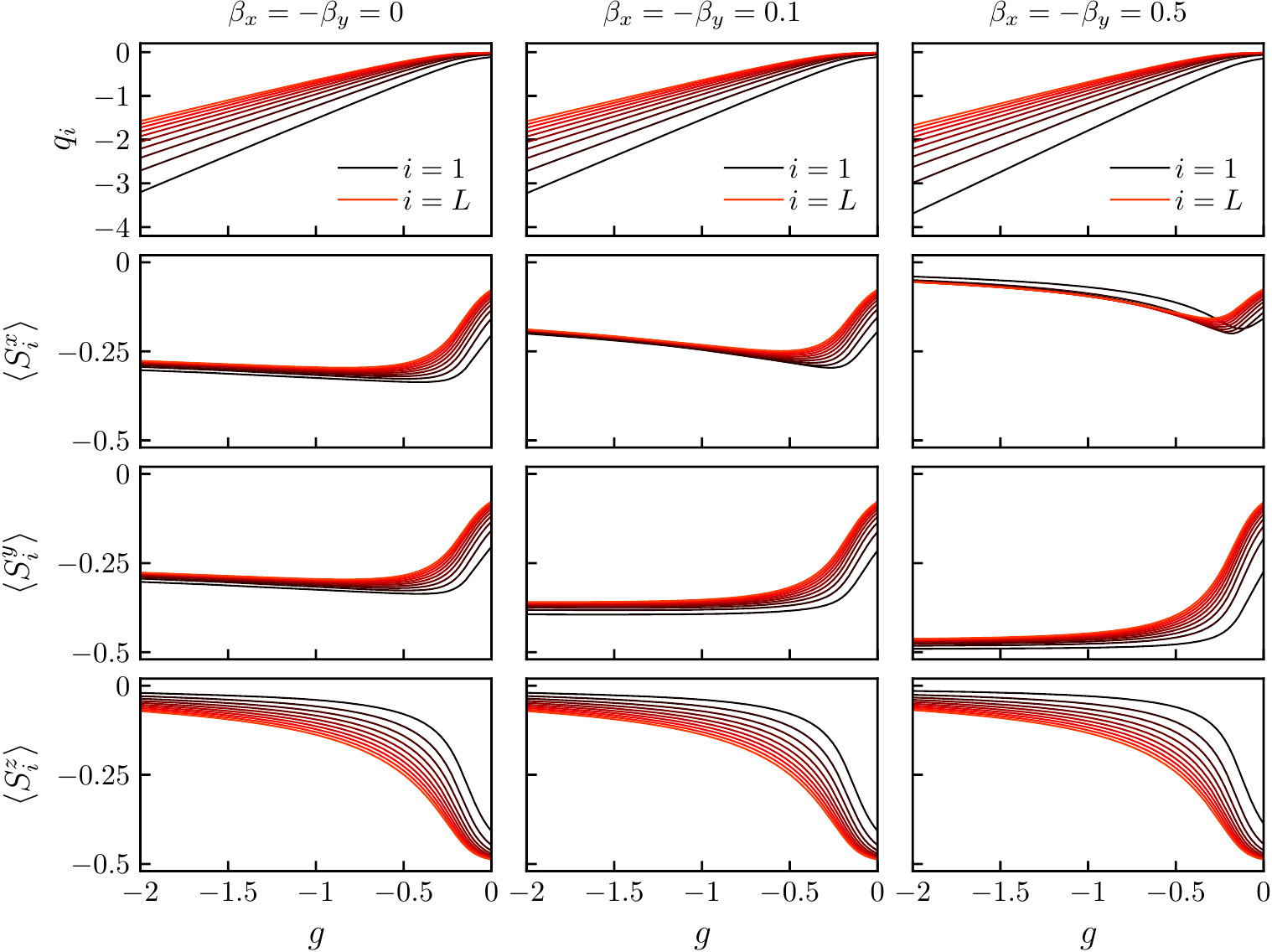}
\caption{Eigenvalues $q_i$ of the conserved charges $Q_i$ and expection values of $\{S_i^x,S_i^y,S_i^z\}$, $\forall i$ for the state connected to $| {\downarrow \dots \downarrow} \rangle$ in the $g=\gamma=\lambda=0$ limit at different values of  $g$. The system is chosen such that $\epsilon_i=i, \forall i$, $\alpha_x=\alpha_y=1$ and $\gamma=\lambda=0.5$ with $L=10$, where $\beta_x=\beta_y=0$ returns the XXZ model and $\beta_x \neq \beta_y$ returns a fully anisotropic model. \label{fig:XYZmodels}}
\end{figure}

We present numerical results for $q_i = \langle Q_i \rangle$ and $\langle S_i^{\alpha} \rangle$ in Fig. \ref{fig:XYZmodels}, where we study the properties of a single given eigenstate which corresponds to the continuous deformation, varying $g$, of the state which would, if no $B_x,B_y$ field was present, correspond to $| {\downarrow \dots \downarrow} \rangle$ in the non-interacting limit $g=0$. As can be seen, at $g=0$, the state corresponds to the lowest projection $-1/2$ of each individual spin along the local orientation of each $\vec{B}_i$, which differs from the fully polarized state along the $z$-axis due the finite $\gamma$ and $\lambda$. The parameters of the system have been chosen in such a way that $\beta_x \neq \beta_y$ returns a fully anisotropic model and $\beta_x=\beta_y=0$ returns the XXZ model. The latter limit returns the known behaviour for the $z$-component of the spin and satisfies $\langle S_i^x \rangle = \langle S_i^y\rangle$ because of the isotropy in the XY-plane. Introducing a finite anisotropy (here included by setting $\beta_x=-\beta_y \neq 0$) breaks this symmetry, as can be observed in the second and third columns. While a detailed analysis is reserved for later works, this Figure clearly shows that for $\beta_x > 0$ and $\beta_y <0$ the expectation values of $S_i^z$ remain largely unaffected, whereas $S_i^y$ moves towards its extremal value of $-0.5$ and $S_i^x$ undergoes an inversion, where the levels with the largest values of $\langle S_i^x \rangle$ at small $g$ have the smallest values of $\langle S_i^x \rangle$ at larger $g$ and vice versa. While these results were presented for a small system with $L=10$, it is possible to go to system sizes of a few hundreds.

\section{Conclusions}
\label{sec:concl}

It was shown how the restriction to spin-$\frac{1}{2}$ Richardson-Gaudin models removes the usual antisymmetry condition on the coupling constants and allows for an explicit solution to the XYZ integrability constraints in the presence of an external magnetic field. After providing a complete parametrisation of these models, it was shown how the resulting set of commuting conserved charges satisfies a set of quadratic equations which can be efficiently solved numerically in order to obtain the eigenvalues. Expectation values of local observables follow from the Hellmann-Feynman theorem, which was used to obtain numerical results highlighting the effect of an anisotropy in the XY plane. The new models can then be used to investigate decoherence effects in central spin systems in more realistic environments containing quadrupole couplings.

\section*{Acknowledgments}
P.W.C. gratefully acknowledges a Francqui Foundation Fellowship from the Belgian American Educational Foundation.

\appendix

\section{Commutators of the conserved charges}
\label{app:commutators}

From direct calculation using only the $su(2)$ commutation relations, it follows that
\begin{eqnarray}
[Q_i,Q_j]&=&i \sum_{\alpha\beta\gamma}\epsilon_{\alpha\beta\gamma}(B_i^{\alpha}\Gamma_{ji}^{\beta}+B_j^{\alpha}\Gamma_{ij}^{\gamma})S_i^{\gamma}S_j^{\beta} \nonumber\\
&+& i \sum_{\alpha\beta\gamma}\sum_{k \neq i,j}\epsilon_{\alpha\beta\gamma}(-\Gamma_{ik}^{\gamma}\Gamma_{ji}^{\beta}-\Gamma_{jk}^{\gamma}\Gamma_{ij}^{\alpha}+\Gamma_{ik}^{\alpha}\Gamma_{jk}^{\beta})S_i^{\alpha}S_j^{\beta}S_k^{\gamma} \nonumber \\
&+& i\sum_{\alpha\beta\gamma}\epsilon_{\alpha\beta\gamma}\Gamma_{ij}^{\alpha}\Gamma_{ji}^{\beta}S_j^{\alpha}S_j^{\beta}S_i^{\gamma} 
-i\sum_{\alpha\beta\gamma}\epsilon_{\alpha\beta\gamma}\Gamma_{ji}^{\alpha}\Gamma_{ij}^{\beta}S_i^{\alpha}S_i^{\beta}S_j^{\gamma}. 
\end{eqnarray}
Demanding the right-hand term to vanish then leads to the usual (antisymmetric) integrability constraints, with the two terms in the third line directly responsible for the antisymmetry constraint. However, making use of the spin-$\frac{1}{2}$ relation $S_i^{\alpha}S_i^{\beta}=\frac{i}{2}\sum_{\gamma}\epsilon_{\alpha\beta\gamma}S_i^{\gamma}+\frac{1}{4}\delta_{\alpha\beta} \mathbbm{1}$ in the last two terms, this can be rewritten as
\begin{eqnarray}
[Q_i,Q_j]&=&i \sum_{\alpha\beta\gamma}\epsilon_{\alpha\beta\gamma}(B_i^{\alpha}\Gamma_{ji}^{\beta}+B_j^{\alpha}\Gamma_{ij}^{\gamma})S_i^{\gamma}S_j^{\beta} \nonumber\\
&+& i \sum_{\alpha\beta\gamma}\sum_{k \neq i,j}\epsilon_{\alpha\beta\gamma}(-\Gamma_{ik}^{\gamma}\Gamma_{ji}^{\beta}-\Gamma_{jk}^{\gamma}\Gamma_{ij}^{\alpha}+\Gamma_{ik}^{\alpha}\Gamma_{jk}^{\beta})S_i^{\alpha}S_j^{\beta}S_k^{\gamma} \nonumber\\
&-&\frac{1}{2} \sum_{\alpha\beta\gamma\kappa}\epsilon_{\alpha\beta\gamma}\epsilon_{\alpha\beta\kappa}\Gamma_{ij}^{\alpha}\Gamma_{ji}^{\beta}S_j^{\kappa}S_i^{\gamma} 
+\frac{1}{2}\sum_{\alpha\beta\gamma\kappa} \epsilon_{\alpha\beta\gamma}\epsilon_{\alpha\beta\kappa}\Gamma_{ji}^{\alpha}\Gamma_{ij}^{\beta}S_i^{\kappa}S_j^{\gamma}.
\end{eqnarray}
Remarkably, the last two terms now mutually cancel after exchanging some dummy-indices, leading to the integrability constraints given in the main text.

\section{Solving the integrability constraints}
\label{app:constraints}

We wish to obtain a solution to the following set of integrability constraints for any permutation $(\alpha, \beta, \gamma)$ of $(x,y,z)$,
\begin{eqnarray}
\Gamma^{\beta}_{i j} B^\alpha_j  + \Gamma^{ \gamma}_{j i} B^\alpha_i = 0,  \qquad \forall  i \neq j, \nonumber\\
\Gamma^{\alpha}_{ik}\Gamma^{\beta}_{jk} - \Gamma^{\alpha}_{ij}\Gamma^{\gamma}_{jk} - \Gamma^{\beta}_{ji}\Gamma^{\gamma}_{ik} = 0, \qquad \forall j \neq k \neq l.
\label{App:Sol:GaudinEq}
\end{eqnarray}
As mentioned in the main text, we can set $B_i^z=1, \forall i$, leading to ($\forall i \neq j \neq k$)
\begin{eqnarray}
\Gamma^{x}_{i j} + \Gamma^{ y}_{j i} =  0, \qquad \Gamma^{y}_{i j} B^x_j  + \Gamma^{z}_{j i} B^x_i =  0, \qquad \Gamma^{x}_{i j} B^y_j  + \Gamma^{z}_{j i} B^y_i = 0, \nonumber \\
\Gamma^{\alpha}_{ik}\Gamma^{\beta}_{jk} - \Gamma^{\alpha}_{ij}\Gamma^{\gamma}_{jk} - \Gamma^{\beta}_{ji}\Gamma^{\gamma}_{ik} = 0.
\end{eqnarray} 
The last two equations in the first line can be combined to return $B^x_i \Gamma^{x}_{ ij}  B^y_j =  - B^x_j \Gamma^{x}_{ji} B^y_i   \equiv \Gamma_{ij}$. Here, we defined a function $\Gamma_{ij}$ which is antisymmetric by construction $\Gamma_{ij}=-\Gamma_{ji}$. Plugging this into the first set of original equations returns (note the asymmetry of $\Gamma^z_{ij}$)
\begin{eqnarray}
\Gamma^{x}_{i j} = \frac{\Gamma_{ij}}{B^x_i B^y_j}, \ \ \ \Gamma^{y}_{ij} = \frac{\Gamma_{ij}}{B^y_i B^x_j},  \ \ \ \Gamma^{z}_{ij} = \frac{\Gamma_{ij}}{B^x_j B^y_j }.
\end{eqnarray}
These satisfy the first line of Eq. \ref{App:Sol:GaudinEq} by construction, and the second line returns
\begin{eqnarray}
\Gamma_{ji}\Gamma_{ik}+\Gamma_{ij}\Gamma_{jk}+\Gamma_{ik}\Gamma_{kj} = 0, \nonumber\\
 \Gamma_{ji}(B^x_i)^{-2}\Gamma_{ik}+\Gamma_{ij}(B^x_j)^{-2} \Gamma_{jk}+ \Gamma_{ik}(B^x_k)^{-2}\Gamma_{kj} = 0, \nonumber\\
\Gamma_{ji}(B^y_i)^{-2} \Gamma_{ik}+\Gamma_{ij}(B^y_j)^{-2} \Gamma_{jk}+\Gamma_{ik}(B^y_k)^{-2} \Gamma_{kj} = 0.
\end{eqnarray}
The first equation is equivalent to an isotropic Gaudin model, whose general solution is given by Gaudin's rational solution $\Gamma_{ij} = \frac{g\gamma\lambda}{\epsilon_i-\epsilon_j}$ (the proportionality factor can be chosen freely and is here chosen as $g \gamma\lambda$). This then allows one to write the two remaining conditions as
\begin{eqnarray}
\frac{(B^x_i)^{-2} -(B^x_j)^{-2} }{\epsilon_{i}-\epsilon_j} = \frac{(B^x_i)^{-2} -(B^x_k)^{-2} }{\epsilon_{i}-\epsilon_k}, \qquad \forall j,k \neq i,
\nonumber\\
\frac{(B^y_i)^{-2} -(B^y_j)^{-2} }{\epsilon_{i}-\epsilon_j} = \frac{(B^y_i)^{-2} -(B^y_k)^{-2} }{\epsilon_{i}-\epsilon_k}, \qquad \forall j,k \neq i.
\end{eqnarray}
The crux of the derivation is that these expressions are independent of the indices $j,k \neq i$ and can only depend on $i$, leading to the general solution 
\begin{eqnarray}
1/(B^x_i)^2  \propto \alpha_x \epsilon_i + \beta_x , \ \ \ \ 1/(B^y_i)^2 \propto  \alpha_y \epsilon_i +\beta_y,
\end{eqnarray}
for arbitrary constants $\alpha_x,\alpha_y,\beta_x,\beta_y$ and proportionality factors which can be chosen as $\gamma^{-2}$ and $\lambda^{-2}$ respectively in order to return the parametrisation in the main text. Keeping $B^z_i$ general at the start of the derivation would have resulted in a similar expression in terms of $\alpha_z \epsilon_i+\beta_z$ and interactions $\Gamma_{ij}^{\alpha}$ also dependent on $\alpha_z$ and $\beta_z$, but it should be clear that this can always be reabsorbed in the other parameters through a rescaling of the conserved charges.

\section*{References}
\bibliographystyle{ieeetr}
\bibliography{MyLibrary.bib}

\begin{thebibliography}{10}




\bibitem{gaudin_1976}
M.~Gaudin, J. Phys. {\bf 37},  1087-1098 (1976)

\bibitem{gaudin_bethe_2014}
M.~Gaudin, {\em {T}he {Bethe} {Wavefunction}}, 
Cambridge University Press (2014)

\bibitem{dukelsky_colloquium_2004}
J.~Dukelsky, S.~Pittel, and G.~Sierra,
{Rev. Mod.  Phys.} {\bf 76}, 643--662 (2004)

\bibitem{ortiz_exactly-solvable_2005}
G.~Ortiz, R.~Somma, J.~Dukelsky, and S.~Rombouts,
{Nucl. Phys. B} {\bf 707}, 421--457 (2005)

\bibitem{lukyanenko_boundaries_2014}
I.~Lukyanenko, P.~S. Isaac, and J.~Links, 
{Nucl. Phys. B} {\bf 886}, 364--398 (2014)

\bibitem{lukyanenko_integrable_2016}
I.~Lukyanenko, P.~S. Isaac, and J.~Links,
{J. Phys. A: Math. Theor.} {\bf 49}, 084001 (2016)

\bibitem{links_solution_2017}
J.~Links, {Nucl. Phys. B} {\bf 916}, 117--131 (2017)

\bibitem{claeys_readgreen_2016}
P.~W.~Claeys, S.~De~Baerdemacker, D.~Van Neck, Phys. Rev. B {\bf 93}, 220503 (2016)

\bibitem{faribault_field_2017}
A.~Faribault and H.~Tschirhart, SciPost Phys. {\bf 3}, 009 (2017) 


\bibitem{sinitsyn_quad_2012}
N. A. Sinitsyn, Yan Li, S. A. Crooker, A. Saxena, and D. L. Smith, Phys. Rev. Lett. {\bf 109}, 166605 (2012)

\bibitem{skrypnyk_integrable_2006}
T.~Skrypnyk, J. Geom. Phys. {\bf 57} 53–67 (2006)

\bibitem{skrypnyk_generalized_2007}
T.~Skrypnyk, {J. Phys. A: Math. Theor.} {\bf 40}, 13337--13352 (2007)

\bibitem{skrypnyk_generalized_2007_1}
T. Skrypnyk, J. Math. Phys. {\bf 48} 113521 (2007)

\bibitem{skrypnyk_non-skew-symmetric_2009}
T.~Skrypnyk, {J. Phys. A: Math. Theor.} {\bf 42}, 472004 (2009)

\bibitem{skrypnyk_non-skew-symmetric_2009-1}
T.~Skrypnyk, {J. Math. Phys.} {\bf 50}, 033504 (2009)

\bibitem{skrypnyk_elliptic_2012}
T.~Skrypnyk, {Nucl. Phys. B} {\bf 856}, 552--576 (2012)

\bibitem{skrypnyk_generalized_2014}
T. Skrypnyk, J. Geom. Phys. {\bf 80} 71–87 (2014)

\bibitem{skrypnyk_twisted_2016}
T.~V. Skrypnyk, {Theor. Math. Phys.} {\bf 189}, 1509--1527 (2016)

\bibitem{skrypnyk_z_2016}
T.~Skrypnyk, {J. Phys. A: Math. Theor.} {\bf 49}, 365201 (2016)

\bibitem{skrypnyk_separation_2017}
T.~Skrypnyk, {J. Phys. A: Math. Theor.} {\bf 50}, 325206 (2017)


\bibitem{babelon_bethe_2007}
O.~Babelon and D.~Talalaev, J. Stat. Mech., P06013 (2007)

\bibitem{faribault_xxxsolve_2011}
A. Faribault, O. El Araby,  C. Str\"ater and V. Gritsev, Phys. Rev. B {\bf 83}, 235124 (2011)

\bibitem{elaraby_xxxsolve_2012}
O.~El Araby, V.~Gritsev and A.~Faribault, Phys. Rev. B {\bf 85}, 115130 (2012)

\bibitem{claeys_xxzsolve_2015}
P. W. Claeys, S. De Baerdemacker, M. Van Raemdonck and D. Van Neck, Phys. Rev. B {\bf 91}, 155102 (2015)

\bibitem{links_completeness_2017}
J.~Links, SciPost Phys. {\bf 3}, 007 (2017)

\bibitem{dimo_quadratic_2018}
C.~Dimo and A.~Faribault, {J. Phys. A: Math. Theor.} {\bf 51}, 325202 (2018)

\bibitem{esebbag_elliptic_2015}
C.~Esebbag and J.~Dukelsky, {J. Phys. A: Math. Theor.} {\bf 48}, 475303 (2015)

\bibitem{relano_excited-state_2016}
A.~Rela\~{n}o, C.~Esebbag, and J.~Dukelsky, {Phys. Rev. E} {\bf 94}, 052110 (2016)

\bibitem{sklyanin_xyz_2016}
E. K. Sklyanin and T. Takebe, Phys. Lett. A {\bf 219}, 217-225 (1996)

\bibitem{dukelsky_class_2001}
J.~Dukelsky, C.~Esebbag, and P.~Schuck, {Phys. Rev. Lett.} {\bf 87}, 066403 (2001)

\bibitem{amico_integrable_2001}
L.~Amico, A.~Di~Lorenzo, and A.~Osterloh, {Phys. Rev. Lett.} {\bf 86}, 5759--5762 (2001)

\bibitem{richardson_restricted_1963}
R.~W. Richardson, {Phys. Lett.} {\bf 3}, 277--279 (1963)

\bibitem{richardson_exact_1964}
R.~W. Richardson and N.~Sherman, {Nucl. Phys.} {\bf 52}, 221--238 (1964)

\bibitem{cambiaggio_integrability_1997}
M.~C. Cambiaggio, A.~M.~F. Rivas, and M.~Saraceno, {Nucl. Phys. A} {\bf 624}, 157--167 (1997)

\bibitem{ibanez_exactly_2009}
M.~Iba{\~n}ez, J.~Links, G.~Sierra, and S.-Y. Zhao, {Phys. Rev. B} {\bf 79}, 180501 (2009)

\bibitem{rombouts_quantum_2010}
S.~M.~A. Rombouts, J.~Dukelsky, and G.~Ortiz, {Phys. Rev. B} {\bf 82}, 224510 (2010)

\bibitem{links_exact_2015}
J.~Links, I.~Marquette, and A.~Moghaddam, {J. Phys. A: Math. Theor.} {\bf 48}, 374001 (2015)

\bibitem{faribault_cs_2013}
A.~Faribault and D.~Schuricht, Phys. Rev. Lett. {\bf 110}, 040405 (2013)

\bibitem{faribault_cs2_2013}
A.~Faribault and D.~Schuricht, Phys. Rev. B {\bf 88}, 085323 (2013)

\bibitem{van_den_berg_competing_2014}
R.~van den Berg, G.~P.~Brandino, O.~El Araby, R.~M.~Konik, V.~Gritsev, and J.-S.~Caux, Phys. Rev. B {\bf 90}, 155117 (2014)

\bibitem{claeys_floquet_2018}
P.~W.~Claeys, S.~De~Baerdemacker, O.~El~Araby, J.-S.~Caux, Phys. Rev. Lett. {\bf 121}, 080401 (2018)

\bibitem{tschirhart_dicke_2018}
H.~Tschirhart, T.~Platini and A.~Faribault, J. Stat. Mech. 083102 (2018) 

\bibitem{rubio_benchmarking_2018}
A. Rubio-García, D.~R.~Alcoba, P.~Capuzzi and J.~Dukelsky, J. Chem. Theory Comput. {\bf 14}, 4183–4192 (2018)

\bibitem{claeys_inner_2017}
P.~W.~Claeys, D.~Van~Neck and S.~De~Baerdemacker, SciPost Phys. {\bf 3}, 028 (2017)

\bibitem{faribault_det_2012}
A.~Faribault and D.~Schuricht, J. Phys. A: Math. Theor. {\bf 45}, 485202 (2012)

\bibitem{tschirhart_det_2014}
H.~Tschirhart and A.~Faribault, J. Phys. A: Math. Theor. {\bf 47}, 405204 (2014)

\bibitem{claeys_det_2015}
P. W. Claeys, S. De Baerdemacker, M. Van Raemdonck and D. Van Neck, J. Phys. A: Math. Theor. {\bf 48}, 425201 (2015)

\bibitem{shen_ground-state_2018}
Y.~Shen, P.~S. Isaac, and J.~Links, {Nucl. Phys. B}, doi:10.1016/j.nuclphysb.2018.08.015 (2018)

\end{thebibliography}

\end{document}